%% file: main.tex
\documentclass[submission, Proceedings]{SciPost}

\usepackage{upgreek}
\usepackage{caption}
\usepackage{subcaption}
\usepackage{xcolor}

\usepackage{lineno}

\hypersetup{
    colorlinks,
    linkcolor={red!50!black},
    citecolor={blue!50!black},
    urlcolor={blue!80!black}
}

\usepackage[bitstream-charter]{mathdesign}
\DeclareSymbolFont{usualmathcal}{OMS}{cmsy}{m}{n}
\DeclareSymbolFontAlphabet{\mathcal}{usualmathcal}

\begin{document}

\begin{center}{\Large \textbf{
DELight: a Direct search Experiment for Light dark matter with superfluid helium\\
}}\end{center}

\input{authorlist}



\definecolor{palegray}{gray}{0.95}
\begin{center}
\colorbox{palegray}{
  \begin{tabular}{rr}
  \begin{minipage}{0.1\textwidth}
    \includegraphics[width=30mm]{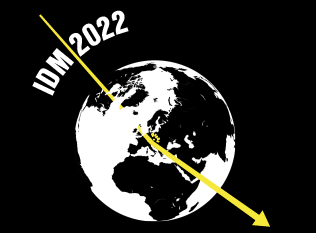}
  \end{minipage}
  &
  \begin{minipage}{0.85\textwidth}
    \begin{center}
    {\it 14th International Conference on Identification of Dark Matter}\\
    {\it Vienna, Austria, 18-22 July 2022} \\
    \doi{10.21468/SciPostPhysProc.?}\\
    \end{center}
  \end{minipage}
\end{tabular}
}
\end{center}

\section*{Abstract}
{\bf
To reach ultra-low detection thresholds necessary to probe unprecedentedly low Dark Matter masses, target material alternatives and novel detector designs are essential. One such target material is superfluid $^4$He which has the potential to probe so far uncharted light Dark Matter parameter space at sub-GeV masses. The new ``Direct search Experiment for Light dark matter'', DELight, will be using superfluid helium as active target, instrumented with magnetic micro-calorimeters. It is being designed to reach sensitivity to masses well below 100\,MeV in Dark Matter-nucleus scattering interactions.}


\section{Introduction}
\label{sec:intro}
Dark Matter (DM) is a well-established concept in particle physics, astrophysics and cosmology and a key parameter in the Lambda Cold Dark Matter (${\rm \Lambda}$CDM) model of Big Bang cosmology. 
A fit of the ${\rm \Lambda}$CDM model to the power spectrum of the cosmic microwave background (CMB) anisotropies predicts that DM constitutes about 85\% of all matter in the Universe \cite{Planck:2018vyg}. Still, although there is compelling evidence for DM \cite{Corbelli_2000,Turner:1990qb,Massey_2010,Planck:2018vyg}, its nature is unknown and DM particles have yet to be discovered. A great variety of theoretical DM particle candidates exists, spanning many orders of magnitude in mass and coupling strength \cite{US_CosmicVision}. Ongoing and planned experiments are only sensitive to a subset of these candidates, one of which being the so-called weakly-interacting massive particle (WIMP) with mass and coupling(s) around the weak scale, produced in the early Universe in thermal equilibrium. Experimental efforts to detect WIMPs include direct detection experiments, designed to measure WIMP scattering off nuclei in the laboratory. Those searches have not discovered the WIMP to date, but have excluded most simple WIMP models with DM masses close to the weak scale \cite{US_CosmicVision}. One possible explanation is that DM is lighter than predicted by the standard WIMP paradigm and well below masses of a few GeV, a possibility that is currently the subject of much theoretical and experimental interest \cite{Lin:2019uvt}. This WIMP-like sub-GeV DM particle candidate is commonly referred to as Light DM (LDM) and its potential couplings to standard matter have been barely probed by direct detection experiments thus far. The signal resulting from DM scattering off nuclei at such low masses is too small to be observed in typical direct DM search experiments and new detector concepts have to be developed to gain the necessary sensitivity. The {\it Direct search Experiment for Light DM}, DELight, will be built to thoroughly explore the LDM region well below the GeV mass scale in DM-nucleus scattering searches.

\section{Superfluid Helium}
\label{sec:helium}



The DELight detector concept exploits the superfluid phase of the nobel gas $^4$He which has many attractive features as target material for a LDM search experiment. It is very light compared to typical direct detection target elements like xenon, argon, germanium and silicon, naturally providing sensitivity to lower DM masses due to better kinematic matching between the DM particle and the target nucleus. Liquid helium is furthermore easily scalable, inexpensive and standard commercial methods exist for its handling. It has no long-lived radioisotopes of its own, it is self-cleaning at superfluid temperatures in that all other atomic species freeze out, and it has a high impedance to external vibration noise. The first excited state of atomic helium is at 19.82\,eV, 
omitting all backgrounds from electronic excitation below this energy \cite{NIST_ASD}.  
Background events near the helium-to-cell interface induced by radioactivity of the surrounding material can be mitigated by fiducializing the monolithic superfluid helium target. Superfluid helium thus provides an extremely radiopure and compact low-background target with means to suppress various external background sources.

A key feature for efficient event classification is the presence of three independent and distinguishable signal channels and the fact that the energy partitioning among these channels depends on the ionization density resulting from the initial particle interaction. The initial interaction prompts a cascade of processes eventually terminating with the total energy distributed among 1) phonons and rotons (collectively referred to as “quasiparticles”), 2) infrared (IR) and especially ultraviolet (UV) photons, and 3) long-lived triplet helium excimers. Quantum evaporation enables the detection of the quasiparticles via liberation of $^4$He atoms into a vacuum \cite{Wya84}. When the quasiparticles are radiated from the interaction site they propagate ballistically at typical speeds of 150-200\,m/sec. Those reaching the free surface of the liquid with enough energy to overcome the helium binding energy of $\sim 0.7$\,meV lead to the evaporation of $^4$He atoms with an efficiency of about 30\,\% \cite{Ban93,Ens94}. The typical quasiparticle energies involved are $\geq 0.8$\,meV. Thus particle interactions with recoil energies of $\mathcal{O}$(10\,keV) yield a large evaporation burst. The UV photons and triplet excimers result from the production of excimers in each event. Excimers in the triplet state are long-lived with a half-life of about 13\,s and are observable as ballistic molecules. Excimers in the singlet state decay within about 1\,ns emitting UV photons with a broad distribution peaking at 16\,eV. Since the first excited state of $^4$He is at $\sim 20$\,eV the liquid is transparent to these UV photons. It was demonstrated in Ref.~\cite{Car17} that the singlet and triplet excimer signals can be separated and that single 15\,eV photons can be detected. 

Because of the favorable properties of superfluid $^4$He it has been considered early on a unique target material for LDM searches \cite{Lan88,Ada96} and remains a target of interest to date as for example in the HeRALD project using superfluid $^4$He instrumented with transition edge sensors (TESs) \cite{Her19}. Careful studies have been carried out to simulate possible backgrounds and to explore DM sensitivities \cite{Ito13,Guo13,Sch16,Her19} and advanced detection schemes involving field ionization have been proposed 
\cite{Mar17}. All these studies underscore the vast opportunities of such a detector. 

\section{Detector Concept}
\label{sec:concept}

\begin{figure}[b!]
     \centering
     \begin{subfigure}[b]{0.48\textwidth}
         \centering
         \includegraphics[width=\textwidth]{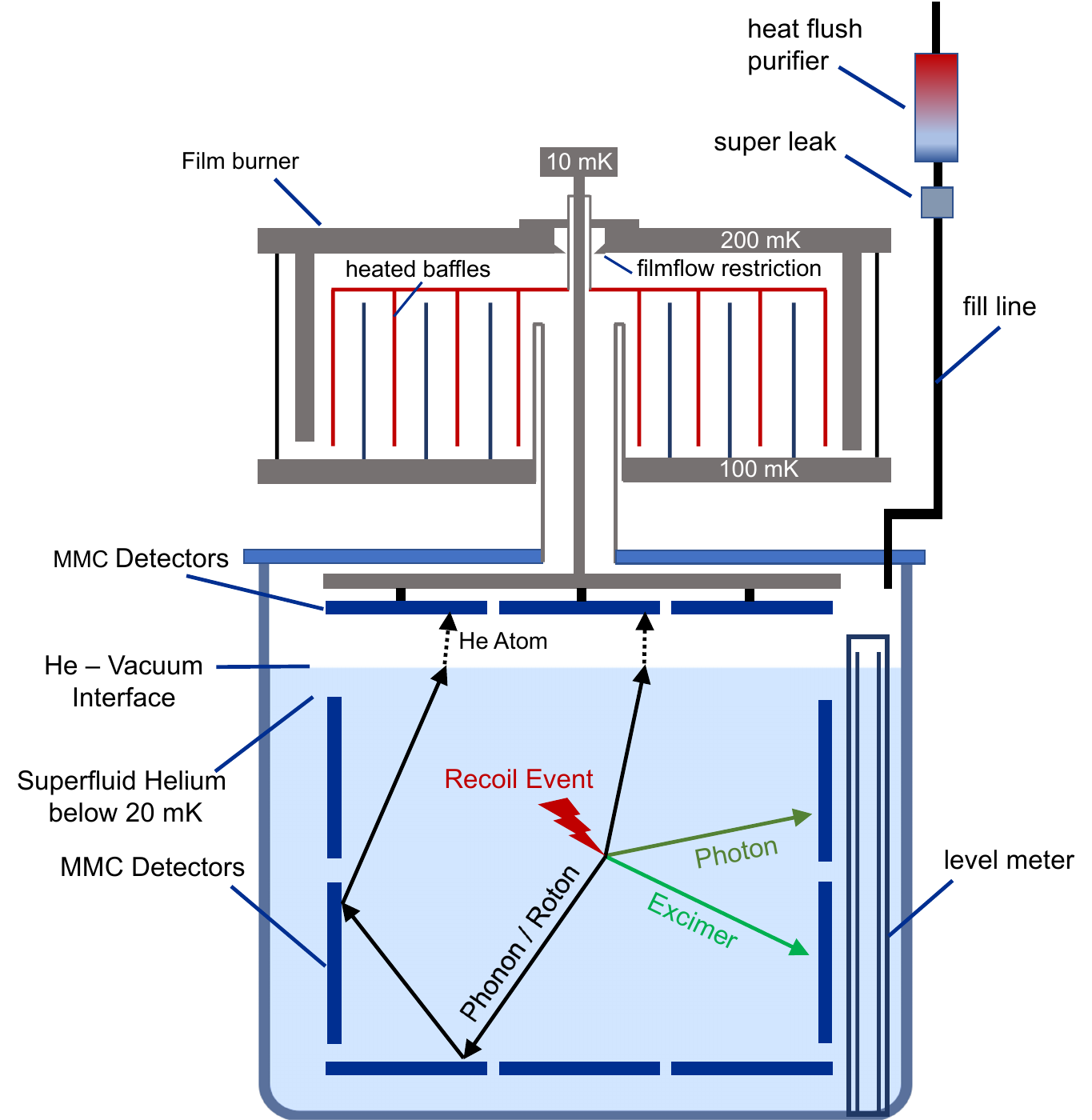}
         \caption{}
         \label{fig:detector}
     \end{subfigure}
     \hfill
     \begin{subfigure}[b]{0.45\textwidth}
         \centering
         \includegraphics[width=\textwidth]{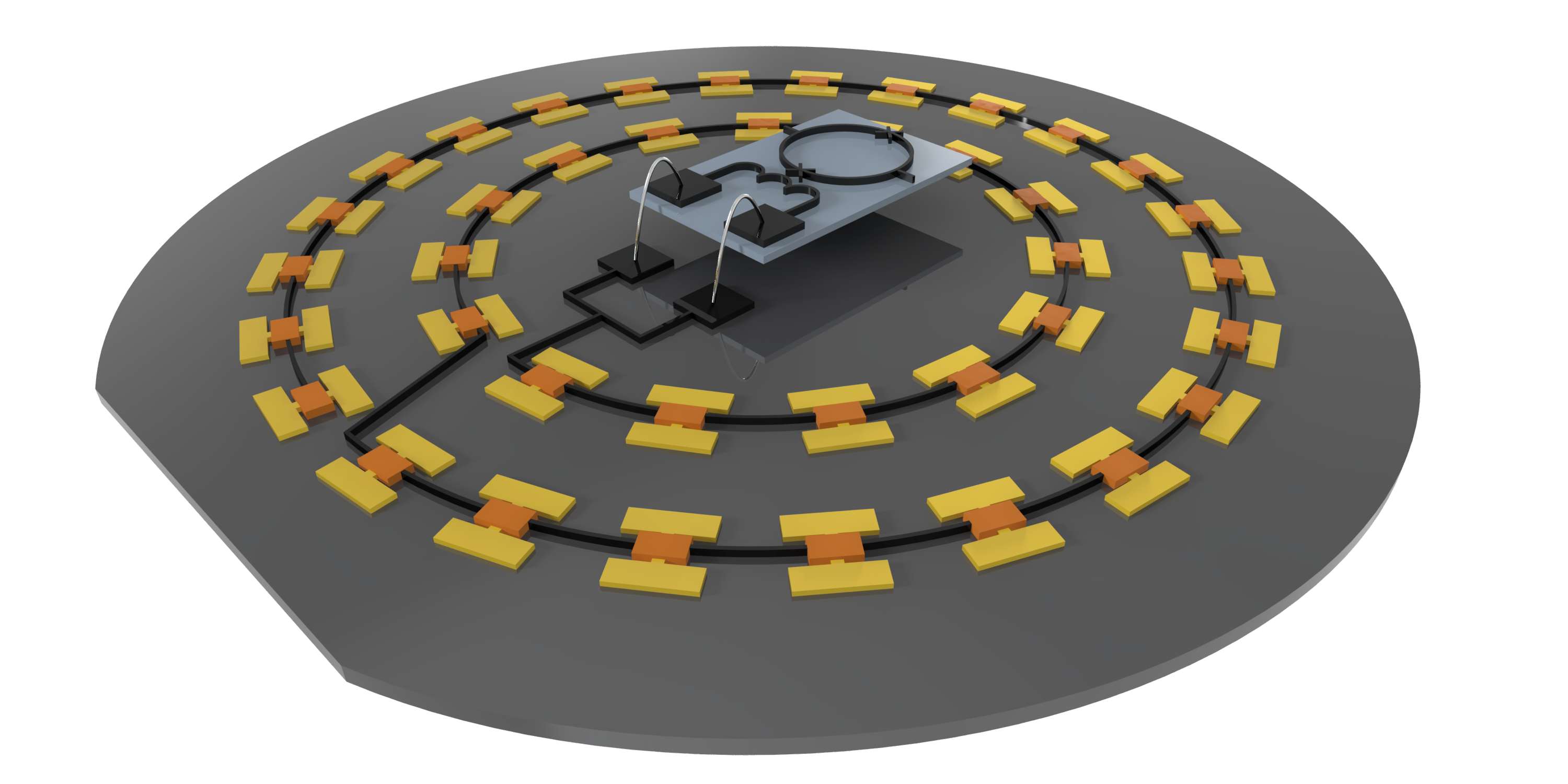}
         \caption{}
         \label{fig:sub-detector}
     \end{subfigure}
        \caption{{\bf a) } Conceptual draft of a DELight helium cell with film burner, heat flush purifier, level sensor and magnetic micro-calorimeter (MMC) sub-detector system. Also illustrated are the main signal channels. {\bf b) } Schematic of a large-area MMC based sub-detector with dielectric wafer as absorber (gray), phonon collectors (yellow), paramagnetic temperature sensor (orange) and superconducting pickup coil (black) connected to a separately mounted readout SQUID \cite{Gray:2016rnt}.}
        \label{fig:schematics}
\end{figure}

\noindent{}The basic principles of a particle detector based on superfluid helium were already developed and demonstrated in the 1990s within the solar neutrino project HERON \cite{Lan87}. The DELight project follows the same basic idea and concept. For HERON a detector was built with a superfluid $^4$He volume of 3\,L, operated at 20$\,$mK. Using movable radioactive sources the underlying detection scheme was established including the quasiparticle generation followed by the liberation of helium atoms through quantum evaporation \cite{Ban93,Ens94}. The evaporated atoms are subsequently adsorbed onto a thin silicon wafer positioned directly above the liquid surface. The adsorption energy of a $^4$He atom onto silicon is about 10 times larger than the binding energy of that atom on the liquid helium surface which provides an effective signal amplification by an approximate factor of 10. As long as helium remains superfluid, though, it creeps up walls including that of the detector cell until it eventually reaches the wafer. To maintain the amplification factor and to reduce the overall heat capacity of the calorimeters, the wafer surface must be kept free from helium which is achievable with a film burner. The film burner is a helium film removal device using heated baffles, as demonstrated by the HERON collaboration \cite{Tor92}. Figure~\ref{fig:detector} shows a draft schematic of the planned DELight detector together with the signal channels described in Sec.~\ref{sec:helium}. The first DELight detector cell will hold a $^4$He volume of 10\,L. In later phases of the experiment larger cell volumes up to $\mathcal{O}$(100\,L) are anticipated.


$^4$He remains liquid down to zero kelvin, enabling energy measurements with ultra-sensitive cryogenic calorimeters. DELight will employ magnetic micro-calorimeters (MMCs) and will be operated below 20\,mK to reduce quasiparticle scattering within the liquid and to enhance the sensitivity of the MMCs. 
The principal components of an MMC are a particle absorber and a paramagnetic temperature sensor \cite{Ban93b,Fle05}. 
The absorber is in tight thermal contact with the sensor and its material is matched to the particles to be observed. 
The sensor is placed in a weak magnetic field to create a temperature dependent sensor magnetization. The temperature and magnetic field dependence 
and the total heat capacity of the micro-calorimeter can be calculated and thus optimized using simulations. 
A change of sensor magnetization resulting from a particle-induced energy deposition, and thus from a rise in temperature, can be measured very precisely as a change of magnetic flux using a superconducting quantum interference device (SQUID). 
The outstanding performance of MMCs was demonstrated in Ref.~\cite{Kem18}, where 6\,keV X-rays were detected with an energy resolution of 1.6\,eV. To keep the MMC in a well-defined state in the absence of an energy deposition, the sensor is weakly linked to a thermal bath with constant temperature. Fundamentally, the energy resolution of MMCs is limited by thermal fluctuations between the absorber, the sensor, and the thermal bath, which are very small at the typical operating temperature of less than 20\,mK \cite{Fle01,Kem18}.


The DELight detector will be instrumented with large-area MMC based sub-detectors, each consisting of a dielectric handling wafer that also acts as absorber, of phonon collectors, of a distributed paramagnetic temperature sensor, and of a superconducting pickup coil that is connected to a readout SQUID. A schematic of such a sub-detector is shown in Fig.~\ref{fig:sub-detector}. Each flux change in one of the pickup coils is transferred to the SQUID and converted into a voltage signal. At the same time, the pickup coils are used to generate the magnetic field required to bias the paramagnetic sensors. The detection system for the first phase of DELight consists of about 50 large-area sub-detectors, each having an area of about 40\,cm$^2$. One fifth of the sub-detectors are placed above the liquid, for the detection of the evaporated He atoms and of UV photons, and four fifths are submerged into the superfluid to also collect UV photons and to additionally observe the long-living triplet excimers (see Fig.~\ref{fig:detector}). Using Si wafers with a thickness of 300\,$\upmu$m and operated at 10\,mK, a detector intrinsic energy resolution of about 3-6\,eV can be estimated \cite{Kem18}, corresponding to a threshold of about 10-20\,eV. Ultimately, at later phases of the experiment, a threshold below 10\,eV is expected to be achieved by optimizing the sub-detector design and the instrumentation layout of the entire detector.

\section{Science Goals}
\label{sec:science}

\begin{figure}[h]
\centering
\includegraphics[width=0.6\textwidth]{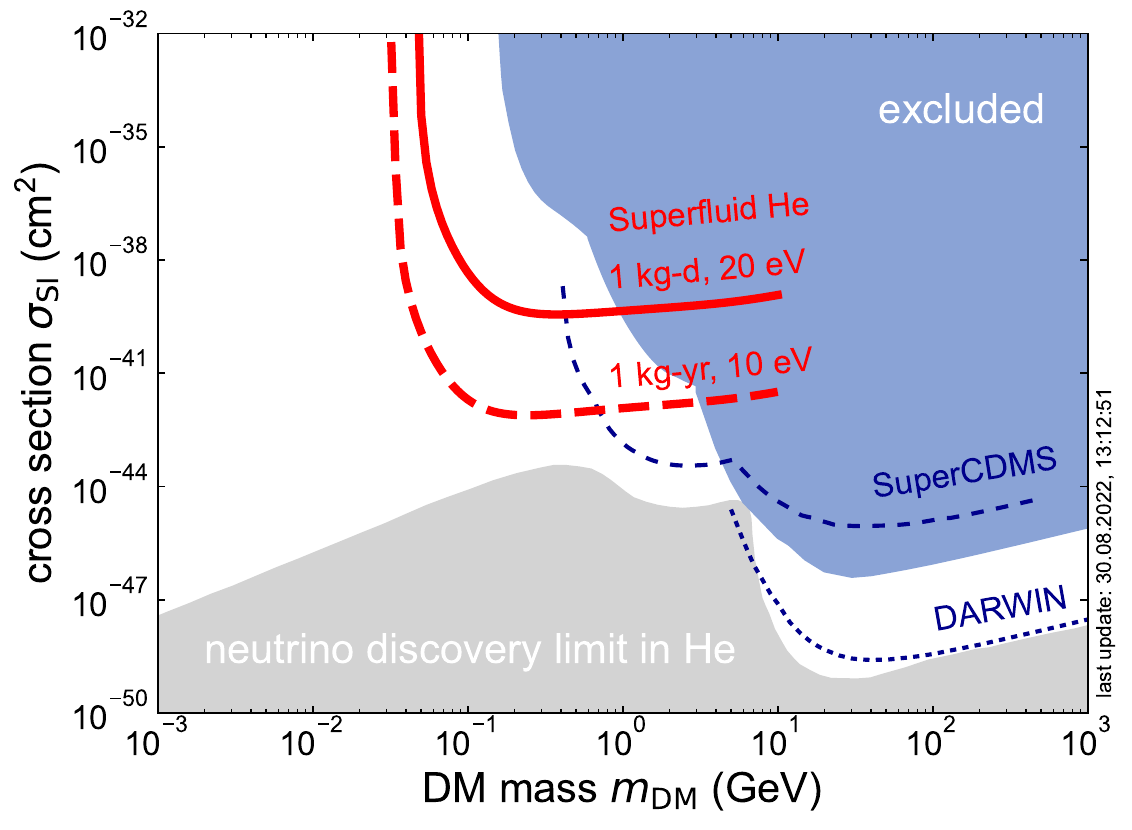}
\caption{Projected spin-independent DM-nucleon scattering limits at 90\% confidence level assuming zero background and the minimum DELight goals in terms of exposure and threshold for phase-I (red solid line) and long-term (red dashed line). Also shown are the neutrino discovery limit in helium calculated as described in Ref.~\cite{Ruppin:2014bra} (gray area), the parameter space excluded by CRESST-III \cite{PhysRevD.100.102002}, DarkSide \cite{Agnes_2018}, and XENON1T \cite{Aprile_2018,Aprile_2021} (blue area) and the projected limits by \mbox{DARWIN} \cite{Aalbers_2016} and \mbox{SuperCDMS} \cite{Agnese_2017} (blue dashed lines).}
\label{fig:science}
\end{figure}

The current status of spin-independent DM-nucleus scattering searches spanning a wide mass range is indicated in Fig.~\ref{fig:science}. At high DM masses at the GeV to TeV scale, DARWIN is foreseen as the ultimate detector based on xenon which will reach the neutrino fog \cite{Aalbers_2016}. Towards lower masses, existing constraints extend to approximately 200\,MeV, but at a substantial loss of sensitivity on the interaction strength. Improvements on the sensitivity to masses around 1\,GeV are expected especially from upcoming solid-state detectors such as CRESST-III and SuperCDMS \cite{PhysRevD.100.102002,Agnese_2017}. Key limitations for an extension towards lower masses are low signal amplitudes expected for massive target nuclei and the dark-current level originating largely from the application of high external fields. The DELight experiment will surpass these limitations by using one of the lightest elements and by not relying on an electric field in the baseline design. Already with a small exposure of $\mathcal{O}$(kg$\cdot$day) and a moderate threshold of about 20\,eV, as planned during the first phase of DELight, unprecedented sensitivity on the scattering cross section is possible at masses below about 1\,GeV based on zero-background projections (see Fig.~\ref{fig:science}). This is supported by \mbox{HeRALD} projections calculated under background assumptions that also qualify for DELight \cite{Her19}. The long range plan of DELight targets a $\mathcal{O}$(kg$\cdot$year) exposure acquired with a helium volume of up to 200\,L in an underground laboratory. With this exposure and an anticipated threshold of $<10$\,eV an LDM mass as low as 30\,MeV becomes accessible. Exploring the sub-100\,MeV mass range forms a milestone in direct DM-nucleus scattering searches.

\section{Conclusion}
\label{sec:conclusion}

Despite great advancement in direct DM searches, DM remains elusive to date using traditional detector materials. Most simple WIMP models have been excluded which motivates searches for models beyond the traditional WIMP. Future projects must thus not only focus on enhancing the detector sensitivity to weaker coupling strengths but also on enlarging the parameter space towards LDM masses well below the GeV scale. A very promising target material for direct LDM-nucleus scattering searches is superfluid $^4$He. It is a very light, ultra-pure, easily scalable, inexpensive target that allows for energy measurements with extremely sensitive micro-calorimeters. The DELight experiment will use superfluid helium as target material instrumented with MMC sensors and a SQUID readout system. DELight is currently in its planning phase and will be designed to probe LDM down to sub-100\,MeV masses. It will become a leading experiment in direct searches for LDM in nuclear scattering interactions.

\section*{Acknowledgements}
We are grateful to G.~Drexlin, R.~Engel, T.~M\"uller, and M.~Weber for their interest in the project and their support and we thank D.~McKinsey and G.~M.~Seidel for fruitful discussions. 

\paragraph{Funding information}
The work of B.v.K. is supported by the Deutsche Forschungsgemeinschaft (DFG) 
under grant No.~420484612. The work of K.V. is supported by the Helmholtz Association through grant No.~W2/W3-118. 
The work of C.E. has received funding by European Union’s Horizon 2020 Research and Innovation Programme, under Grant Agreement No.~824109.



\bibliography{main.bib}

\nolinenumbers

\end{document}

%% file: authorlist.tex
\begin{center}
B.~von~Krosigk\textsuperscript{1,2 $\star$},
K.~Eitel\textsuperscript{1},
C.~Enss\textsuperscript{2,3},
T.~Ferber\textsuperscript{4},
L.~Gastaldo\textsuperscript{2},
F.~Kahlhoefer\textsuperscript{5},
S.~Kempf\textsuperscript{6,3},
M.~Klute\textsuperscript{4},
S.~Lindemann\textsuperscript{7},
M.~Schumann\textsuperscript{7},
F.~Toschi\textsuperscript{1,7} and
K.~Valerius\textsuperscript{1} 
\end{center}

\begin{center}
{\bf 1} Institute for Astroparticle Physics (IAP), Karlsruher Institut f\"ur Technologie (KIT), 76344 Eggenstein-Leopoldshafen, Germany
\\
{\bf 2} Kirchhoff-Institute for Physics, Heidelberg University, 69120 Heidelberg, Germany
\\
{\bf 3} Institute for Data Processing and Electronics (IPE), Karlsruher Institut f\"ur Technologie (KIT), 76344 Eggenstein-Leopoldshafen, Germany
\\
{\bf 4} Institute of Experimental Particle Physics (ETP),  Karlsruher Institut f\"ur Technologie (KIT), 76131 Karlsruhe, Germany
\\
{\bf 5} Institute for Theoretical Particle Physics (TTP), Karlsruher Institut f\"ur Technologie (KIT), 76128 Karlsruhe, Germany
\\
{\bf 6} Institute of Micro- and Nanoelectronic Systems (IMS), Karlsruher Institut f\"ur Technologie (KIT), 76187 Karlsruhe, Germany
\\
{\bf 7} Institute of Physics, Freiburg University, 79104 Freiburg, Germany
\\
* bkrosigk@kip.uni-heidelberg.de
\end{center}